\documentclass{article}
\usepackage {amssymb}
\begin {document}
\author {Albert Schwarz
\\University of California at Davis,\\Davis,CA 95616 
 \thanks{Partially
supported by NSF
grant DMS-9801009}}
\title{Noncommutative supergeometry and duality.}

\maketitle
\large
\begin {abstract}
\large
We introduce a notion of $Q$-algebra that can be considered as a
generalization of the notion of $Q$-manifold (a supermanifold equipped
with an odd vector field obeying $\{ Q,Q\} =0$). We develop the theory of
connections on modules over $Q$-algebras and prove a general duality
theorem for gauge theories on such modules. This theorem contains as a
simplest case $SO(d,d,{\bf Z})$-duality of gauge theories on
noncommutative tori.
\end {abstract} 
  It was shown recently that  noncommutative geometry is quite useful in
the study of string theory/M-theory (see [4]-[10] and references therein).
It
was proved ,in particular,
that the gauge theory on noncommutative tori has $SO(d,d,{\bf Z})$ duality
group, closely related to T-duality in string theory [5]. The goal of
present paper is to derive a very general duality theorem, containing
$SO(d,d,{\bf Z})$-duality as a special case. This theorem is formulated
and proved in the framework of  "noncommutative supergeometry". The main
idea of   noncommutative geometry is to consider every associative algebra
as an algebra of functions on  "noncommutative space".  Of  course,
supergeometry fits very nicely in this approach: one of the most
convenient definitions of a supermanifold is formulated in terms of the
algebra of functions on it.  One can say that supergeometry is
"supercommutative ${\bf Z}_2$-graded  noncommutative geometry".  

   One of important notions of supergeometry is the notion of $Q$-manifold
(of a manifold equipped with an odd vector field $Q$ satisfying $\{ Q,Q\}
=0$);see [12]. The first order differential operator $\hat {Q}$
corresponding to
$Q$ obeys $\hat {Q}^2=0$; therefore the algebra of functions on a
$Q$-manifold can be considered as a differential ${\bf Z}_2$-graded
associative algebra  and it is naturally to think that differential ${\bf
Z}_2$-graded associative algebras are analogs of $Q$-manifolds. However,
in this paper we introduce another notion, the notion of $Q$-algebra, that
also can be considered as a natural generalization of $Q$-manifold and
that can be used to develop the theory of connections and to prove a
general duality theorem. Namely, we define a $Q$-algebra as a ${\bf
Z}_2$-graded associative algebra   $\Omega$ equipped with an odd
derivation
$Q$ obeying $Q^2a=[\omega ,a]$; here $\omega \in \Omega$ should satisfy
$Q\omega =0$.(One says that a linear operator acting on graded algebra is
a derivation, if it satisfies the graded Leibniz rule.) 
Of course, in the case when $\Omega$ is supercommutative
this definition coincides with the definition of differential algebra, but
if we do not assume  supercommutativity this definition is more general.
We define a connection on $\Omega$-module ${\cal E}$   as a map $\nabla
:{\cal E}\rightarrow {\cal E} $  obeying the Leibniz rule $\nabla
(ea)=\nabla e\cdot a +(-1)^{\deg e}e\cdot Qa$; the general duality theorem
is formulated in terms of such connections. We analyze the relation of the
standard definition of connection in  noncommutative  geometry to our one.
It seems that many well known constructions and theorems become more
transparent in our formalism. From the other side many considerations of
present paper are similar to arguments employed previously, especially in
[1], [2], [3]. 

\centerline {{\bf Preliminaries.}}
When we talk about  associative algebra  $\Omega$ we always have in mind
graded (${\bf Z}$-graded or ${\bf Z}_2$-graded) unital associative algebra
over ${\bf C}$.  Graded commutator is defined by the formula 
$$[a,b]=ab-(-1)^{\deg a\cdot \deg b}ba.$$
In what follows all commutators are understood as graded commutators.

   A (right) module ${\cal E}$ over ${\Omega}$ is a graded vector space
with operator of multiplication on elements of $\Omega$ from the right;
this operation should have standard properties: $(ea)\cdot b=e\cdot (ab)\
\  e(a+b)=ea+eb$ etc. Grading on ${\cal E}$ should be compatible with
grading on $\Omega$ (i.e. $\deg (ea)=\deg e+\deg a$). The definition of a
left module is similar; by default our modules are right modules. 

  A vector space ${\cal E}$ is called an $(\Omega_1,\Omega _2)$-bimodule
if it is a left $\Omega_1$-module and a right $\Omega _2$-module; we
require that $(a_1e)a_2=a_1(ea_2)$ where $a_i\in \Omega _i,\ \  e\in {\cal
E}$.

  If  ${\cal E}_1,{\cal E}_2$ are $\Omega$-modules we define an
$\Omega$-homomorphism as a map $\varphi: {\cal E}_1\rightarrow {\cal E}_2$
obeying  $\varphi (xa)=\varphi (x)a$.  The graded algebra of
$\Omega$-homomorphisms of the  $\Omega$-module ${\cal E}$ into itself
(algebra of $\Omega$-endomorphisms) is denoted by End$_{\Omega}{\cal E}$.
If ${\cal E}$ is an $(\Omega_1,\Omega _2)$-bimodule there exist natural
homomorphisms  $\Omega _1\rightarrow {\rm End}_{\Omega _2}{\cal E}$  and
$\Omega _2\rightarrow {\rm End}_{\Omega _1}{\cal E}$.

  If  ${\cal E}_1$ is a right $\Omega$-module and ${\cal E}_2$ is a left
$\Omega$-module we define ${\cal E}_1\otimes _{\Omega}{\cal E}_2$ as a
vector space obtained from the standard tensor product ${\cal E}_1\otimes
_{{\bf C}}{\cal E}_2$ by means of identification $e_1a\otimes e_2\sim
e_1\otimes ae_2$, where $e_i\in {\cal E}_i,\ \  a\in \Omega$.

    We say that  $(\Omega_1,\Omega _2)$-bimodule $P_1$ and
$(\Omega_2,\Omega _1)$-bimodule $P_2$ are Morita equivalence bimodules if
$P_1 \otimes _{\Omega _2}P_2$ is isomorphic to $\Omega _1$  as
$(\Omega_1,\Omega _1)$-bimodule  and $P_2\otimes _{\Omega _1} P_1$ is
isomorphic to $\Omega _2$ as   $(\Omega_2,\Omega _2)$-bimodule. If such
bimodules $P_1,P_2$ exist we say that $\Omega _1$ and $\Omega _2$ are
Morita equivalent. For every $\Omega _1$-module we can define an $\Omega
_2$-module $\tilde {{\cal E}}={\cal E}\otimes _{\Omega _1}P_1$.  Applying
an
analogous construction to $\Omega _2$-module ${\tilde {\cal E}}$ and
bimodule $P_2$ we obtain
an inverse map: ${\cal E}=\tilde {{\cal E}}\otimes _{\Omega_2}P_2$. One
can say that Morita equivalent algebras have equivalent categories of
modules.

      A  (finitely generated) free module $\Omega  ^n$ over $\Omega$ can
be defined as the space of column vector with entries from $\Omega$ and
with componentwise  multiplication on elements of $\Omega$. We regard
$\Omega ^n$ as a right module, but it can be considered also as
$(\Omega,\Omega )$-bimodule.  (We already used the structure of
$(\Omega,\Omega )$-bimodule on $\Omega ^1=\Omega$.) The algebra
End$_{\Omega}\Omega^n$ of endomorphisms of $\Omega ^n$ can be identified
the algebra Mat$_n{\Omega}$  of  $n\times n$ matrices with entries from
$\Omega$; these matrices act on $\Omega ^n$ by means of multiplication
from the left. A projective $\Omega$-module can be defined as a direct
summand $E$ in a free module $\Omega ^n$.  The decomposition $\Omega
^n=E+E^{\prime}$ into a direct sum determines an endomorphism $e: \Omega
^n\rightarrow \Omega ^n$ projecting  $\Omega ^n$ onto $E$; in other words
$e^2=e,\ \  ex=x$ for $x\in E,\ \  ex^{\prime}=0$ for $x^{\prime}\in
E^{\prime}$.  Notice that in our  terminology projective modules are
always finitely generated.

  Projective $\Omega$-modules form a semigroup with respect to direct
summation. Applying the Grothendieck construction to this semigroup we
obtain the K-theory group $K_0(\Omega)$. More precisely, we say that a
projective module $E$ specifies an element $[E]\in K_0(\Omega)$  and
impose the relation $[E _1+E_2]=[E_1]+[E_2]$.  A ${\bf C}$-linear map
$\tau : \Omega\rightarrow {\bf C}$ is called a (graded ) trace if it
vanishes on all (graded) commutators: $\tau ([a,b])=0$ for all
$a,b\in\Omega$. We always consider graded traces; therefore we almost
always omit the word "graded"  in our formulations. 

  A trace $\tau$ on  $ \Omega $ generates a trace on End$_{\Omega}\Omega
^n=$Mat$_n\Omega$; this trace will be denoted by the same symbol $\tau$.
(To calculate the trace of a matrix $(a_{ij})\in$Mat$_n\Omega$ one should
take the supertrace of the matrix $(\tau(a_{ij}))$. 

  If  $E\subset \Omega ^n$ is a projective module then the algebra
End$_{\Omega}E$ of endomorphisms of $E$ can be identified with the
subalgebra of End$_{\Omega}\Omega ^n=$Mat$_n\Omega$ consisting of elements
of the form $eae$. (Here $e:\Omega ^n\rightarrow \Omega ^n$ is a
projection of $\Omega^n$ onto $E,\ \  a\in $End$_{\Omega}\Omega ^n$). We
define a graded trace $\tilde {\tau}$ on  End$_{\Omega}E$ as a restriction
of $\tau$ to this subalgebra. 

     If ${\cal E}$ is an $\Omega$-module, then starting with an element
$g\in \Omega $ and $\Omega$-homomorphism $f: {\cal E}\rightarrow \Omega$
we can construct an endomorphism $g\otimes f:{\cal E}\rightarrow {\cal E}$
transforming $x\in {\cal E}$ into $gf(x)\in {\cal E}$. (The endomorphism
$g\otimes f$ can be considered as a generalization of linear operator of
rank $1$.)

   For any algebra $\Omega $ we construct a vector space $\bar {\Omega}=
\Omega /[ \Omega , \Omega]$ factorizing the vector space $ \Omega $ with
respect to the subspace $[ \Omega , \Omega ]$ spanned by all (graded)
commutators $[a,b]$.  This construction is closely related with the notion
of trace: traces on  $\Omega$  correspond to linear functionals on  $\bar
{\Omega }$.  

   If ${\cal E}$ is a projective $\Omega $-module, one can construct a
${\bf C}$-linear map Tr: End$_{\Omega}{\cal E}\rightarrow \bar {\Omega}$
transforming an endomorphism of the form $g\otimes f$ into the class
$\overline {f(g)} \in \bar {\Omega}$ of  $f(g)\in {\Omega}$. (Such a ${\bf
C}$-linear map is unique because in the case of projective module every
endomorphism can be represented as a finite sum of endomorphisms of the
form $g\otimes f$.) The map Tr has the main property of trace 
$${\rm Tr} [\varphi ,\psi ]=0$$
(trace of graded commutator of two $\Omega$-endomorphisms $\varphi,\psi
\in {\rm End}_{\Omega}{\cal E}$ vanishes).  In some sense the map
Tr:End$_{\Omega}{\cal E}\rightarrow \bar {\Omega}$ can be considered as
universal trace on End$_{\Omega}{\cal E}$. (As we mentioned every trace
$\tau$ on $\Omega$ determines a trace $\tilde {\tau}$ on
End$_{\Omega}{\cal E}$. It is easy to verify that $\tilde
{\tau}(\varphi)=\tau({\rm Tr }\varphi)$.)

\centerline {\bf $Q$-algebras}
Definition. {\it  Let $\Omega$ be a graded associative algebra. We say the
$\Omega$ is a $Q$-algebra if it is equipped with derivation $Q$ of degree
$1$ and there exists an element $\omega \in \Omega ^2$ satisfying 
\begin {equation}
Q ^2x=[\omega ,x]
\end {equation}
for all $x\in \Omega$.}

  Calculating $Q^3x$ in two ways we obtain 
\begin {equation}
Q^3x=Q([\omega,x])=Q\omega\cdot x +\omega\cdot Qx-Qx\cdot \omega -x\cdot
Q\omega\cdot (-1)^{\deg x}
\end {equation}
\begin {equation}
Q^3x =Q^2\cdot Qx=[\omega, Qx]=\omega\cdot Qx-Qx\cdot \omega.
\end {equation}
We see that $Q\omega\cdot x=x\cdot Q\omega\cdot (-1)^{\ deg x}$, i.e. 
$$[Q\omega,x]=0$$
We proved that $Q\omega\in \Omega ^3$ commutes with all elements of
$\Omega $ (in the sense of superalgebra). In almost all interesting cases
it follows from this condition that $Q\omega $ vanishes. 

{\it We will include the additional condition 
$$Q\omega =0$$
in the definition of $Q$-algebra.}

We always consider unital algebras. It is easy to to check that applying
$Q$ to the unit we get 0. (This follows from the  Leibniz rule.)

   Let us consider a (graded) $\Omega$-module $E$. We define a connection
on ${\cal E}$ as a ${\bf  C}$-linear operator $\nabla :E\rightarrow {\cal
E}$ having degree $1$ and obeying the Leibniz rule:
\begin {equation}
\nabla (xa)=(\nabla x)\cdot a+(-1)^{\deg x}\cdot x\cdot Qa.
\end {equation}
for all $x\in E,\ \ a\in \Omega$.

  Let us introduce the notation 
$$ \hat {a}x=(-1)^{\deg x\cdot \deg a}xa$$
The formula (4) can be rewritten in the form 
$$[\nabla,\hat {a}]=\widehat {Qa}$$
It is easy to check that some standard statements about connections remain
true in our case. However, the definition of curvature should be modified.

  1) If  $\nabla $ is a fixed connection on ${\cal E}$, then every other
connection has the form 
$$\nabla ^{\prime}=\nabla +A$$
where $A\in {\rm End}^1_{\Omega}{\cal E}$ is an arbitrary endomorphism of
degree $1$.

  2)  If $\varphi \in{\rm End}_{\Omega}{\cal E}$ is an endomorphism then
$[\nabla ,\varphi ]$ is also endomorphism. 

  3) The operator $\nabla ^2+\hat {\omega}$ is an endomorphism: $\nabla
^2+\hat {\omega} \in {\rm End}^2_{\Omega}{\cal E}$. This endomorphism is
called the curvature of connection $\nabla$; it is denoted by $F(\nabla)$
(or simply by $F$). It obeys $[\nabla, F]=0$. 

  To check this statement we represent $\nabla ^2$ as ${1\over
2}[\nabla,\nabla]$ and calculate $[[\nabla,\nabla],\hat {a}]$ by means of
(4) and Jacobi identify.

  4) Let us define the operator $\tilde {Q}: {\rm End}_{\Omega}{\cal
E}\rightarrow {\rm End}_{\Omega}{\cal E}$ by the formula 
$$ \tilde {Q} \varphi =[\nabla,\varphi].$$
It is easy to verify that 
\begin {equation}
 \tilde {Q}^2\varphi =[F,\varphi],
\end {equation}
where $F$ is the curvature of $\nabla$.

It follows from this statement and from $\tilde {Q}F=[\nabla,F]=0$ that
the algebra ${\rm End}_{\Omega}{\cal E}$ equipped with the operator
$\tilde {Q}$ is a $Q$-algebra with $\tilde {\omega}=F$. (One should
notice, however, that we can also take $\tilde {\omega}=F+c$, where $c$ is
a central element obeying $\nabla c=0$.)

  Let us consider an $(\Omega _1,\Omega _2)$-bimodule $P$ where $\Omega
_1$ is a $Q$-algebra with respect to the operator $Q_1$ and $\Omega _2$ is
a $Q$-algebra with respect to the operator $Q_2$. We say that an operator
$\nabla _P:P\rightarrow P$ is a connection on bimodule $P$ if 
$$\nabla _P(ax)=(-1)^{\deg a}\cdot a \nabla _P(x)+Q_1a\cdot x$$
$$\nabla _P(xb)=\nabla_Px\cdot b+(-1)^{\deg x}\cdot x Q_2b$$
for all $x\in P,\ \ a\in \nabla _1,\ \  b\in \nabla_2$.

   In other words, $\nabla _P$ should be a connection with respect to
$\Omega _1$ and with respect to $\Omega _2$ at the same time. 

  It follows from the above statements that every $\Omega$-module ${\cal
E}$ equipped with a connection  $\nabla$ can be considered as $({\rm
End}_{\Omega}{\cal E},\Omega)$-bimodule and $\nabla$ is a connection on
this bimodule.

  Using an $(\Omega _1, \Omega _2)$-bimodule $P$ we can assign to every
(right ) $\Omega _1$-module ${\cal E}$ a (right) $\Omega _2$-module
$\tilde {{\cal E}}$ taking the tensor product with $P$:
\begin {equation}
\tilde {{\cal E}}={\cal E}\otimes _{\Omega _1}P
\end {equation}
(To take the tensor product over $\Omega _1$ we identify $ea\otimes p$
with $e\otimes ap$ in the standard tensor product ${\cal E}\otimes _{\bf
C}P$. Here $e\in {\cal E},\ \  p\in P,\ \  a\in \Omega _1$.)

   If we have a connection $\nabla _P$ in the bimodule $P$ we can transfer
a connection on ${\cal E}$ to a connection on $\tilde {{\cal E}}$. Namely,
for every connection $\nabla $ on ${\cal E}$ we define an operator $\nabla
\otimes 1+1 \otimes \nabla_P$ on ${\cal E}\otimes _{\bf C}P$. It is easy
to check that this operator is compatible with identification $ea\otimes
p\sim e \otimes ap$ and therefore descends to an operator $\tilde
{\nabla}: \tilde {{\cal E}} \rightarrow \tilde {{\cal E}}$. The operator
$\tilde {\nabla}$ can be considered as a connection on $\Omega _2$-module
$\tilde {{\cal E}}$. 
  
   It is easy to relate the curvatures of the connections $\nabla$ and
$\tilde {\nabla}$. We should take into account that correspondence between
${\cal E}$ and $\tilde {{\cal E}}$ is natural, i.e. to every endomorphism
$\sigma \in  {\rm End}_{\Omega_1}{\cal E}$ we can assign an endomorphism
$\tilde {\sigma}\in {\rm End}_{\Omega _2} \tilde {{\cal E}}$ (the map
$\sigma \otimes 1:{\cal E}\otimes _{{\bf C}}P\rightarrow {\cal E}\otimes
_{{\bf C}}P$ descends to an endomorphism $\tilde {\sigma}:\tilde {{\cal
E}}\rightarrow \tilde {{\cal E}}$). In particular, the curvature
$F(\nabla)
\in {\rm End}_{\Omega _1}{\cal E}$ determines an endomorphism $\widetilde
{F(\nabla)} 
\in  {\rm End}_{\Omega _2} \tilde {{\cal E}}$. One can verify that the
curvature $F(\tilde {\nabla})$ of the connection $\tilde {\nabla}$ on
$\tilde {{\cal E}}$ can be represented in the form: 
\begin {equation}
F(\tilde {\nabla})=\widetilde {F(\nabla)}+\tilde {\varphi},
\end {equation}  
where $\tilde {\varphi}$ is a fixed element of  ${\rm End}_{\Omega _2}
\tilde {{\cal E}}$ .

  To verify (7) we notice that 
$$\nabla ^2\otimes 1+\hat {\omega}_1\otimes 1: {\cal E}\otimes _{{\bf
C}}P\rightarrow {\cal E}\otimes _{{\bf C}}P$$ descends to the endomorphism
$\widetilde {F(\nabla)}: \tilde {{\cal E}}\rightarrow \tilde {{\cal E}}$
and $\nabla ^2\otimes 1+1\otimes \nabla^2_P+1\otimes \hat {\omega}_2$
descends to  $F(\tilde {\nabla}): \tilde {{\cal E}}\rightarrow \tilde
{{\cal E}}$. Using the relation $\hat {\omega }_1\otimes 1=-1\otimes
\hat{\omega}_1$ we obtain that the map $\tilde
{\varphi}=F(\tilde{\nabla})-\widetilde{F(\nabla)} $is induced by the map
$\varphi =1\otimes \psi : {\cal E}\otimes _{{\bf  C}}P\rightarrow {\cal
E}\otimes _{{\bf C}}P$ where the map $\psi : P\rightarrow P$ is given by
the formula 
$$\psi =\nabla ^2_P+\hat {\omega}_1+\hat {\omega }_2.$$
It is easy to check  that
\begin {equation}
 \psi  \in {\rm End}_{\Omega _1} P\cap {\rm End}_{\Omega _2} P 
\end {equation}
(i.e. $\psi (ax)=a\psi (x),\ \  \psi (xb)=\psi (x)b$ for $x\in P,\ \  a\in
\Omega _1,\ \  b\in \Omega _2$). To check that $\psi$ commutes with $a\in
\Omega _1$ we represent it in the form $\psi =\widehat
{F_1(\nabla_P)}+\hat {\omega}_2$, where $F_1(\nabla_P)$ stands for the
curvature of $\nabla_P$ considered as $\Omega _1$-connection; the
representation $\psi =\widehat {F_2(\nabla_P)}+\hat {\omega}_1$ should be
used to prove that $\psi \in {\rm End }_{\Omega _2}P $. 

   It follows from (8) that $\varphi =1\otimes \psi $ descends to $\tilde
{{\cal E}}$ and gives an $\Omega_2$-endomorphism $\tilde {\varphi}$. One
should notice that these facts are clear also from the representation
$\tilde {\varphi}=F(\tilde {\nabla})-\widetilde {F(\nabla)}$. 

   To illustrate the above statements we can start with an arbitrary
$Q$-algebra $\Omega$ and arbitrary $\Omega$-module $P$ with connection
$\nabla _P$. We consider $P$ as $(\Omega_1,\Omega_2)$-bimodule, where
$\Omega_1={\rm End }_{\Omega}P,\ \  \Omega_2=\Omega$. (We have seen that
$\Omega _1={\rm End}_{\Omega}P$ is a $Q$-algebra with respect to the
operator $\tilde {Q}\varphi =[\nabla _P,\varphi ]$ and that $\nabla _P$ is
a connection also with respect to this $Q$-algebra.) It follows from our
calculations that $F=F_2(\nabla_P)=\nabla ^2_P+\hat {\omega}_2,\ \  \hat
{\omega }_1=-F$ and therefore $\psi =F+\hat {\omega}_1=0$. (We can obtain
the same result noticing that $F_1(\nabla
_2)=\nabla^2_P+\hat{\omega}_1=(F-\hat {\omega}_1=-\hat {\omega}_2$.) We
see that in our situation $\varphi =0$;  hence, $F(\tilde
{\nabla})=\widetilde {F(\nabla)}$. (However, as we noticed above one can
modify the definition of $Q$-algebra ${\rm End} _{\Omega}P$ adding central
element $c$ with $\nabla c=0$ to $\omega _1$; then $\varphi \not= 0$.) 

   We would like to give conditions when gauge theories on
$\Omega_1$-module ${\cal E} $ and in $\Omega_2$ -module $\tilde {{\cal
E}}$ are equivalent. To establish such an equivalence we need $(\Omega
_1,\Omega _2)$-bimodule $P^{\prime}$ equipped with connection
$\nabla_{P^{\prime}}$. Such a bimodule permits us to transfer modules and
connections in opposite direction. If the constructions obtained by means
of $P^{\prime}$ are inverse to constructions specified by $P$ we say that
bimodules $P,P^{\prime}$ give  Morita equivalence of $Q$-algebras
$\Omega_1$ and $\Omega _2$ (or that they are  Morita equivalence
bimodules). Of course, this notion generalizes the standard notion of
Morita equivalence of associative algebras, when we do not use the
operator $Q$ and connections. The definition of Morita equivalence
bimodules can be reformulated in the following more constructive way. Let
us suppose that there exist two bilinear scalar products between $P$ and
$P^{\prime}$ taking values in $\Omega _1$ and in $\Omega _2$ respectively.
We assume that scalar products are $\Omega _2$-invariant and $\Omega
_1$-invariant correspondingly. In other words, we assume that for $p\in
P,\ \  p^{\prime}\in P^{\prime}$ we have $<p,p^{\prime}>_1\in \Omega_1,\ \
<p^{\prime},p>_2\in \Omega_2$ and $<p\omega,p^{\prime}>_1=<p,\omega
p^{\prime}>_1$ for $\omega \in \Omega _2$,
$<p^{\prime}\sigma_1,p>_2=<p^{\prime},\sigma_1 p>_2$ for $\sigma_1 \in
\Omega_1$. We require also that 
\begin {equation}
\sigma_1 <p,p^{\prime}>_1\sigma_2=<\sigma _1p,p^{\prime}\sigma _2>_1,\ \
\omega_1<p^{\prime},p>_2\omega _2=<\omega_1p^{\prime},p\omega_2>_2
\end {equation}
\begin {equation}
p_1<p,p^{\prime}>_1=<p_1,p>_2p^{\prime},\ \
<p^{\prime},p>_2p_1^{\prime}=p^{\prime}<p,p_1^{\prime}>_1
\end {equation}
Here $p,p_1\in P,\ \  p^{\prime},p_1^{\prime}\in P^{\prime},\ \  \sigma
_i\in \Omega _1,\ \  \omega _i\in \Omega _2$. The scalar products
determine maps 
$$\alpha :P\otimes _{\Omega  _2}P^{\prime}\rightarrow \Omega _1,\ \  \beta
:P^{\prime} \otimes _{\Omega_1} P \rightarrow \Omega _2.$$
We can consider $P\otimes _{\Omega _2}P^{\prime}$ and $\Omega _1$ as 
   $(\Omega _1,\Omega _1)$-bimodules; then it follows from (9), that
$\alpha$ is a homomorphism of bimodules; similarly $\beta$ is a
homomorphism of   $(\Omega _2,\Omega _2)$-bimodules. We require that
$\alpha$ and $\beta$ be isomorphisms. Then 
$$({\cal E}\otimes _{\Omega_1} P)\otimes _{\Omega _2}P^{\prime}={\cal
E}\otimes _{\Omega_1}(P\otimes_{\Omega_2}P^{\prime})={\cal E}\otimes
_{\Omega _1}\Omega _1={\cal E}$$
for every  $\Omega _1$-module ${\cal E}$. This statement together with
similar statement for   $\Omega _2$-modules gives us one-to-one
correspondence between    $\Omega _1$-modules and    $\Omega _2$-modules
(more precisely it gives us equivalence of categories of  $\Omega
_1$-modules and    $\Omega _2$-modules).  To obtain  one-to-one
correspondence between  connections we should impose additional
requirements 
\begin {equation}
<\nabla _Pp,p^{\prime}>_1+<p,\nabla_{P^{\prime}}p^{\prime}>_1=
Q<p, p^{\prime}>_1, 
\end {equation}
$$<\nabla_{P^{\prime}}p^{\prime},p>_2+<p^{\prime},\nabla_{P^{\prime}}p>_2=
Q<p^{\prime}, p>_2$$

It follows from our assumptions that the operator$$\nabla _P \otimes 1 +
1 \otimes \nabla _{P^{\prime}}$$ on $P \otimes _{\bf C} {P^{\prime}}$
descends to operator $Q$ on
$P \otimes _{\Omega_2} {P^{\prime}}$. Using that $Q\cdot  1 =0$  we obtain  
that the operator $\nabla \otimes 1 + 1 \otimes Q$ on
${\cal E} \otimes_{\bf C}\Omega_1$ descends to $\nabla$ on
${\cal E }\otimes_{\Omega_1}\Omega_1={\cal E}$. This means that going from
$\Omega_1$-connections to $\Omega_2$-connections and back we obtain the
original $\Omega_1$-connection. This fact together with similar statement
about $\Omega_2$-connections gives one-to-one correspondence between 
$\Omega_1$-connections and $\Omega_2$-connections.

We see that under our
 conditions we have equivalence between gauge theories on 
  $\Omega _1$-module ${\cal E}$ and on   $\Omega _2$-module $\tilde {{\cal
E}}$ (duality). We will describe later how the duality of gauge theories
on noncommutative tori can be obtained this way. 

Let us study connections on projective $\Omega$-modules where $\Omega$ is
a $Q$-algebra.

   First of all it is easy to construct a connection on an arbitrary
projective $\Omega$-module ${\cal E}$. Namely, if ${\cal E}$ is specified
by means of projection $e:\Omega ^n\rightarrow \Omega ^n$ (i.e. $e\Omega^n
={\cal E}$) we can construct a connection on ${\cal E}$ (so called
Levi-Civita connection)  by means of the formula $D=eQe$ where $Q$ acts on
$\Omega ^n$ componentwise. (The Leibniz rule for $D$ follows from $e^2=e$
and from the Leibniz rule for $Q$.) The curvature of the Levi-Civita
connection is given by the formula: 
$$F=e((Qe)^2+\omega\cdot 1).$$ 
 For any algebra $\Omega$ we defined a vector space
$\bar{\Omega}=\Omega/[\Omega,\Omega]$. If $\Omega$ is a $Q$-algebra we
have $Q([\Omega,\Omega])\subset [\Omega,\Omega]$. This means that the
operator $Q: \Omega\rightarrow \Omega $ descends to an operator $\bar
{Q}:\bar {\Omega}\rightarrow \bar {\Omega}$.  It is easy to check that
$\bar {Q}$ is a differential: $\bar {Q}^2=0$.

   Now we will define the Chern character of a connection $D$ on a
projective $\Omega$-module ${\cal E}$ as an element of $\bar {\Omega}$:
$${\rm ch}D=\sum_{q=0}{1\over q!}{\rm Tr}F^q$$
(Recall that we defined a map ${\rm Tr: End}_{\Omega}{\cal E}\rightarrow
\bar {\Omega}$ using the formula ${\rm Tr} (g\otimes f)=\overline {f(g)}$.
Here $f: {\cal E} \rightarrow \Omega$ is an $\Omega$-homomorphism, $g\in
\Omega$ and $g\otimes f$ transforms $x\in {\cal E}$ into $gf(x)\in {\cal
E}$. The map $a\rightarrow \bar {a}$ transforms $a\in \Omega$ into its
class $\bar {a}\in \bar {\Omega}$.)

   One can prove the following statements: 

1) ${\rm ch}D$ is closed with respect to the differential $\bar {Q}$ in
$\bar {\Omega}$:
\begin {equation}
\bar {Q}{\rm ch}D=0
\end {equation}

2) If $D^{\prime},D$ are two connections on $\Omega$-module ${\cal E}$
then ${\rm ch}D^{\prime}-{\rm ch}D$ is exact with respect to the
differential $\bar {Q}$: 
\begin {equation}
{\rm ch}D^{\prime}-{\rm ch}D=\bar {Q}{\rm (something)}.
\end {equation}
The proof is based on the following lemma:

For every endomorphism $\varphi \in {\rm End}_{\Omega}{\cal E}$ we have 
\begin {equation}
{\rm Tr} [D,\varphi]=\bar {Q}{\rm Tr}\varphi
\end {equation}
It is sufficient to verify (14) for Levi-Civita connection $D=eQe$
(because ${\rm Tr} [D^{\prime}-D,\varphi]=0$) and for endomorphisms of the
form $\varphi =g\otimes f $ (because these endomorphisms span
End$_{\Omega}{\cal E}$).

  Using (14) we deduce (12) from the relation $[D,F^q]=0$ that follows
immediately from $[D,F]=0$.
 
  To derive (13) we will consider a smooth family $D(t)=D+t(D^{\prime}-D)$
of connections on ${\cal E}$ and prove that 
$${d\over dt}{\rm ch}D(t)=\bar{Q}({\rm something}).$$
First of all we notice that the curvature $F(t)$ of connection $D(t)$
obeys 
$${dF(t)\over dt}=[\Gamma,D(t)]$$ 
where $\Gamma =D^{\prime}-D\in {\rm End}_{\Omega}{\cal E}$. We see that 
$${dF\over dt}=[\Gamma,D] \ \ \ {\rm mod} [\bar {\Omega},\bar {\Omega}].$$
and therefore
$${dF^q\over dt}=q[\Gamma,D]F^{q-1}=q[D,\Gamma F^{q-1}] \ \ \  {\rm
mod}[\bar {\Omega},\bar {\Omega}],$$
$${d{\rm Tr}F^q\over dt}=q{\rm Tr}[D,\Gamma F^{q-1}]\in \bar {Q}(\bar
{\Omega}).$$
Integrating over $t$ we obtain (13). 

   In the proof of (13) we assumed that $\Omega $ is equipped with
topology having some properties that permit us to justify the calculations
above. These assumptions are not necessary; it is easy to modify our
consideration  to obtain completely algebraic proof (as in [2] for
example). 

   Sometimes it is convenient to reformulate (13) using the notion of
closed trace. We say that a linear functional on $\Omega$ is a closed
trace if it vanishes on (graded) commutators and on elements of the form
$Qa$.  It follows from (13) that for a closed trace $\tau$ the number
$\tau({\rm ch}(D))$ does not depend on the choice of the connection $D$ on
the module ${\cal E}$; it depends only on the $K$-theory class of the
module ${\cal E}$. 

   Using the differential $\bar {Q}$ we can define the homology $H(\bar
{\Omega})$ in the standard way: $H(\bar {\Omega})=\ker \bar {Q}/{\rm
Im}\bar {Q}$. It follows from (12), (13) that the Chern character
specifies a homomorphism ch: $K_0(\Omega)\rightarrow H^{{\rm
even}}(\bar{\Omega})$. One can construct also a map
$K_1(\Omega)\rightarrow H^{{\rm odd}}(\bar{\Omega})$; we will not discuss
this construction here.

\centerline {\bf  Connections on modules over associative algebras.}

   The theory of connections on modules over $Q$-algebras can be
considered as a generalization of the theory of connections on associative
algebras. If  $A$ is an associative algebra one can construct a
differential ${\bf Z}$-graded algebra $\Omega (A)=\sum _{n\geq 0}\Omega
^n(A)$ (universal differential graded algebra) in the following way. The
vector space $\Omega ^n(A)$ is is spanned by formal expressions
$a_0da_1...da_n$ and $\lambda da_1...da_m$ where $a_0,...a_n\in A,\ \
n\geq 0,\ \  m\geq 1,\ \  \lambda \in {\bf C}$. The multiplication and the
differential on $\Omega (A)$ are defined by means of Leibniz rule and
relation $d^2=0$. If ${\cal E}$ is an $A$-module we define on $\Omega
(A)$-module ${\cal E}^{\prime}$ as a tensor product:${\cal
E}^{\prime}={\cal E}\otimes _A\Omega (A)$ where $\Omega (A)$ is considered
as $(A,\Omega (A))$-bimodule. We can define a connection on $A$-module
${\cal E}$ as a connection of $\Omega (A)$-module ${\cal E}^{\prime}$;
this definition is equivalent to the definition given by Connes (see[1]).

   In this definition of connection on $A$-module ${\cal E}$ the 
algebra $\Omega(A)$ 
can be replaced with any differential extension of the algebra $A$ (with
any
differential graded algebra $\Omega$ that contains $A$ as a subalgebra
of
$\Omega ^0$). Moreover, one can consider any $Q$-extension of $A$ (any
$Q$-algebra $\Omega$ obeying $A\subset \Omega^0$) and define a
connection on $A$-module $E$ as a connection on $\Omega$-module
$E\otimes _A\Omega$. It is interesting to notice that under certain
conditions on algebra $\Omega $ any projective $\Omega$-module ${\cal
E}$ can be represented in the form $E\otimes _A\Omega$ where $E$ is
projective $A$-module, $A=\Omega^0$ (see[11]). In particular, this
statement is correct if $\Omega=\sum_{0\leq k\leq n}\Omega^k$ (i.e.
the 
degree of an element of $\Omega$ is non-negative and bounded from
above). 

   If a Lie algebra $L$ acts on $A$ by means of
infinitesimal automorphisms  (derivations) we can construct a differential
graded algebra $\Omega=\Omega (L,A)$ of cochains of Lie algebra $L$ with
values  in $A$. The elements of $\Omega$ can be considered as $A$-valued
functions of anticommuting variables $c^1,....,c^n$ corresponding to the
elements of the basis $\delta_1,...,\delta_n \in L$;  the differential $d$
has the form 
$$d\omega =(\delta
_{\alpha}\omega)c^{\alpha}+{1\over2}f^{\alpha}_{\beta\gamma}c^{\beta}c^{\gamma}{\partial\over\partial
c^{\alpha}}$$ 
where $f^{\alpha}_{\beta\gamma}$ are the structure constants of $L$ in the
basis $\delta_1,...,\delta _n$. 

  In other words we can describe the vector space $\Omega (L,A)$ as a
tensor product $\Lambda (L^*)\otimes A$ where $L^*$ stands the vector
space dual to $L$ and $\Lambda (M)$ denotes the Grassmann algebra
generated by vector space $M$ (as vector space $\Lambda (M)$ is a direct
sum of antisymmetric tensor powers of $M$). The grading on $\Omega (L,A)$
is defined by means of the natural grading on $\Lambda (L^*)$; if $A$ is a
graded algebra one should take into account also the grading on $A$.

  Let us consider in more detail connections on $A$-module $E$ with
respect
to differential extension $\Omega =\Omega(L,A)$. In this case ${\cal
E}^0=E\otimes _A\Omega^0=E,\ \ {\cal E}^1=E\otimes _A\Omega^1=E\otimes
_{{\bf C}} L^*$. The elements $e\otimes \omega,\ \  e\in E,\ \  \omega\in
\Omega$ span ${\cal E}$, therefore, the connection $\nabla :{\cal
E}^r\rightarrow {\cal E}^{r+1}$ is completely determined by the map
$\nabla :{\cal E}^0\rightarrow {\cal E}^1$ that can be considered as a map
$\nabla :E\rightarrow E\otimes L^*$ or as a family of maps
$\nabla_x:E\rightarrow E$ that depend linearly on $x\in L$. Instead of the
family  $\nabla _x$ we can consider $n$ maps $\nabla _1,...,\nabla_n$
corresponding to the elements of the basis $f_1,...,f_n$ of the Lie
algebra $L$. These maps obey the Leibniz rule 
$$\nabla _{\alpha}(ea)=\nabla _{\alpha}e\cdot a+e\delta _{\alpha}a$$
where $\delta _{\alpha}$ stands for the derivation of the algebra $A$
that corresponds to $f_{\alpha}\in L$.

  Let $A$ be an algebra $A_{\theta}$ of smooth functions on
$d$-dimensional noncommutative torus (i.e. an algebra of expressions of
the form $\sum c_nU_n$, where $c_n$ is a ${\bf C}$-valued function on
a
$d$-dimensional lattice that vanishes at infinity faster than any power
and
the multiplication is defined by the formula $U_nU_m=\exp (\pi
i\theta_{nm})U_{n+m}$, where $\theta _{nm}$ is a bilinear function on the
lattice). Then it is natural to construct a differential extension of
$A_{\theta}$ taking as $L$ the Lie algebra of derivations $\delta _x$
where $\delta _xU_l=<x,l>U_l$. (We assume that the lattice is embedded
into vector space $V$. The vector $x$ belongs to the dual space $V^*$ that
can be identified with the Lie algebra $L$.) Connections corresponding to
this differential extension of $A_{\theta}$ appear naturally in the study 
of toroidal
compactifications of M(atrix) theory. 

  Let us suppose now that in addition to the action of Lie algebra on $A$
we have a finite group $G$ acting on $A$ and $L$ by means of automorphisms
and that actions of $G$ on $A$ and $L$ are compatible. (If we denote
automorphisms of $A$ and of $L$ corresponding to the element $\gamma\in G$
by the same letter $\gamma$ this means that $\gamma (T(a))=(\gamma T)\cdot
(\gamma a)$ for every $\gamma \in G,\ \ T\in L,\ \ a\in A$.) One can
define
in natural way the action of $G$ on the algebra $\Omega =\Omega (L,A)$; 
this action commutes with the differential. This means that we can regard
the crossed product $\Omega\rtimes G$ as a differential algebra; we have
$(\Omega \rtimes G)_0=\Omega _0\rtimes G=A\rtimes G$ and therefore the
crossed product can be considered as differential extension of $A\rtimes
G$. 

  An $A\rtimes G$-module $E$ can be considered as an $A$-module equipped
with action of the group $G$ that is compatible with the action of $G$ on
$A$ (more precisely we should have $\gamma (xa)=\gamma (x)\cdot \gamma
(a)$). As always a connection on $E$ is defined as a connection $\nabla$
on $\Omega \rtimes G$-module
 $${\cal E}=E\otimes _{A\rtimes
G}(\Omega \rtimes G).$$ 
Again this connection is completely
determined by the map $\nabla : {\cal E}^0\rightarrow {\cal
E}^1$ that can be considered as a map
$$\nabla :E\rightarrow E\otimes _{A\rtimes G}(\Omega ^1(L,A)\rtimes G)$$
or as a map
$$\nabla :E\rightarrow E\otimes L^*$$
that determines a connection on $A$-module $E$ and is compatible with the
action of the group $G$ on $E$ and on $E\otimes L^*$.

  In the case when $A$ is an algebra of functions on noncommutative torus
the connections we obtained are precisely the connections that arise by
compactification of M(atrix) theory on toroidal orbifolds 
(see [9],[10]).

\centerline {\bf Conclusion.}
  
  In present paper we generalized the theory of connections on modules
over associative algebra. We embedded this theory into the theory of  
connections on modules over $Q$-algebras and proved a general duality
theorem in this framework. Namely, we proved that under certain conditions
there exists one-to-one correspondence between connections on modules over
one $Q$-algebra and connections on modules over another $Q$-algebra and
found
relation between corresponding curvatures. More precisely, it follows
from our results that  under certain conditions gauge theory constructed
by means of  
$Q$-algebra $\Omega$ is equivalent to gauge theory corresponding to the
$Q$-algebra $\rm End_{\Omega}\cal E$ where $\cal E$ is an
$\Omega$-module equipped with a connection.( We use the fact that every 
connection determines a structure of $Q$-algebra on the algebra of
endomorphisms.)
This theorem can be applied to many concrete
situations; we are planning to give applications in forthcoming papers. 

\centerline {\bf Acknowledgments.}

I am very grateful to M. Khovanov, M. Kontsevich,  M. Rieffel and
D. Sternheimer for useful comments.
I am deeply indebted to Caltech Theory Group, especially to J. 
Schwarz and E. Witten, for warm hospitality and interesting
discussions.
   
{\bf {\centerline  {Appendix}}}

   {\bf {\centerline  {Differential algebras,$ A_\infty$-algebras and
$Q$-algebras.}}}

 If $\Omega$ is a $Q$-algebra one can extend it to a 
 differential algebra $\Omega ^{\prime}$ adjoining new element $X$ obeying
 $X^2=\omega$, $a_1Xa_2=0$ for all $a_1, a_2 \in \Omega$. The  
 differential on $\Omega ^{\prime}$ is defined by the formula $d\omega
 =Q\omega +[X,\omega ]$ (this construction was used by A.Connes [1] ).
 The differential algebra we constructed is equivalent in some sense 
to the $Q$-algebra we started with.

One can define an $A_\infty$-algebra as a vector space $V$ equipped
with multilinear operations $m_i$  ; these operations should satisfy some
relations . ( The operations $m_i$ determine a derivation of tensor
algebra over $V$; the square of this derivation should be equal to zero.)
In standard definition of $A_\infty$-algebra one considers operations
$m_i$ where the number of arguments $i$ is $\geq 1$.
However, one can modify the definition  including an operation $m_0$
(if the number of arguments is equal to zero, then the operation is
simply a fixed element of $V$). Using the modified definition one
can say that $Q$-algebra is an $A_\infty$-algebra where all operations
with the number of arguments $\geq 3$ vanish. ( In standard definition
this requirement leads to differential algebras.)

\centerline {\bf References.}

  1. Connes, A., {\it Noncommutative Geometry}, Academic Press,New York,
(1994).

  2. Karoubi, M., {\it Homology cyclique et K-theorie}, Asterisque, 149
(1987), 147 pp.

  3. Kastler, D., {\it Cyclic cohomology within the differential 
envelope}, Hermann, Paris, (1988),184 pp.

  4. Connes, A., Douglas, M., and Schwarz, A., {\it Noncommutative
geometry and Matrix theory: compactification on tori}, 
JHEP {\bf 02} (1998) 003; hep-th/9711162.

  5. Schwarz, A., {\it Morita equivalence and duality}, Nucl. Phys. {\bf
B534} (1998) pp. 720-738; hep-th/9805034.

  6. Konechny, A. and Schwarz, A., {\it BPS states on noncommutative tori
and duality}, Nucl. Phys. {\bf B550} (1999) 561-584; hep-th/9811159.

   Konechny, A. and Schwarz, A.,{\it Supersymmetry algebra
and BPS states of super Yang-Mills theories on noncommutative tori}, 
Phys. Lett. {\bf B453} (1999) 23-29; hep-th/9901077.

   Konechny, A. and Schwarz, A.,{\it 1/4 BPS states on noncommutative
tori}, JHEP {\bf 09} (1999) 030; hep-th/9907008.

  7. Nekrasov, N. and Schwarz, A., {\it Instantons on noncommutative
$R^4$, and $(2,0)$ superconformal six-dimensional theory}, hep-th/9802068.

  8. Seiberg, N. and Witten, E., {\it String Theory and Noncommutative
Geometry}, 
JHEP {\bf 9909} (1999) 032; hep-th/9908142.

  9. Ho, P.-M. and Wu, Y.-S., {\it Noncommutative
Gauge Theories in Matrix Theory}, 
Phys.Rev. {\bf D58} (1998) 066003; hep-th/9801147.

 10.    Konechny, A. and Schwarz, A.,{\it Compactification of M(atrix)
theory on noncommutative toroidal orbifolds}, hep-th/9912185

 11. Bass, H., {\it Algebraic K-theory}, Benjamin, NY-Amsterdam, (1962),
762 pp

 12. Schwarz, A.
{\it Geometry of 
Batalin-Vilkovisky quantization}, CMP, 
{\bf 155} (1993) 249-260,
{\it Semiclassical approximation in
Batalin-Vilkovisky formalism}, CMP,{\bf 158} (1993) 373-396,
Alexandrov,M., Kontsevich, M., Schwarz, A. and Zaboronsky, O.
,{\it  The geometry of master eqution and topological quantum field
theory},
Int. J. of Modern Physics,{\bf  A12} (1997) 1405-1429 

\end{document}